# Macroscopic quantum tunneling of magnetization explored by quantum-first-order reversal curves (QFORC)

Fanny Béron[1], Miguel A. Novak[2], Maria G. F. Vaz[3], Guilherme P. Guedes[3,4], Marcelo Knobel[1], Amir Caldeira[1], Kleber R. Pirota[1,*]

[1]Instituto de Física Gleb Wataghin, Universidade Estadual de Campinas, 13083-859, Campinas (SP), Brazil

[2]Instituto de Física, Universidade Federal do Rio de Janeiro, 21941-972, Rio de Janeiro (RJ), Brazil

[3]Instituto de Química, Universidade Federal Fluminense, 24210-346, Niterói (RJ), Brazil

[4]Instituto de Ciências Exatas, Departamento de Química, Universidade Federal Rural do Rio de Janeiro, 23890-000, Seropédica (RJ), Brazil.

**Abstract**

A novel method to study the fundamental problem of quantum double well potential systems that display magnetic hysteresis is proposed. The method, coined quantum-first-order reversal curve (QFORC) analysis, is inspired by the conventional first-order reversal curve (FORC) protocol, based on the Preisach model for hysteretic phenomena. We successfully tested the QFORC method in the peculiar hysteresis of the $Mn_{12}Ac$ molecular magnet, which is governed by macroscopic quantum tunneling of magnetization. The QFORC approach allows one to quickly reproduce well the experimental magnetization behavior, and more importantly to acquire information that is difficult to infer from the usual methods based on matrix diagonalization. It is possible to separate the thermal activation and tunneling contributions from the magnetization variation, as well as understand each experimentally observed jump of the magnetization curve and associate them with specific quantum transitions.

The quantum double well potential (QDWP) is one of the most important potential profiles in quantum mechanics, because it admits states which are linear superpositions of quantum mechanical states with 'classical' analogues, an important concept related to quantum computation [1]. One can find examples where the theory of QDWP can be successfully applied in chemistry, biology or physics, including the tunneling of the magnetic flux in superconducting quantum interference devices (SQUIDs), tunneling dynamics of substitutional defects in solids, or hydrogen pair transfer in the hydrogen-bonded cyclic dimers [2]. A remarkable example of QDWP physics is the macroscopic quantum tunneling (MQT) of magnetization in nanomagnets. Indeed, MQT has developed into a subject of great interest after the introduction of the concept by Caldeira and Leggett in the beginning of the 1980's [3, 4]. In this context, single molecule magnets–are model systems that allow the observation of quantum tunneling of the magnetization, thermally assisted quantum tunneling and resonant tunneling of magnetic moment [5, 6]. The first described molecule of this type was the [$Mn_{12}O_{12}(CH_3COO)_{16}(H_2O)_4$] (hereafter $Mn_{12}Ac$) [7]. This molecule is composed of 12 interacting $Mn^{III}$ and $Mn^{IV}$ ions and has a $S = 10$ spin ground state with high uniaxial anisotropy. Well isolated from each other they present a superparamagnetic behavior above a blocking temperature of 3 K [8], and a temperature dependent hysteresis with steps due to thermally assisted resonant quantum tunneling of its magnetization (see fig. 1) [6, 9]. Below 1 K these steps become temperature independent as pure quantum tunneling turns to be the dominant reversal process. Hamiltonian with complex anisotropy contributions combined with the Landau-Zener model [10] has been used to explain the experimental results. There are still some controversies between experimental results and theoretical predictions despite the large amount of results in the literature.

In this letter we present a novel experimental protocol to study the MQT of the magnetization in $Mn_{12}Ac$ single crystals. Named quantum-first-order reversal curve (QFORC), it is inspired by the classical first-order reversal curve (FORC) technique [11]. In the FORC paradigm, based on the classical Preisach model [12], the hysteresis is represented by a set of hysteresis operators, called hysterons, consisting of elementary square hysteresis loops [Fig. 2 (a)]. Here, the elementary hysterons, called quantum hysterons (q-hysterons), represent the respective spin states transitions between the two sides of the potential energy barrier. As a remarkable new result, the QFORC procedure allows deconvoluting the quantum tunneling behavior from thermally activated magnetization reversal procuresses, in function of the applied field. In principle, it can be easily applicable in many other two level quantum systems that present quantum tunneling and hysteresis.

In this context fundamental differences separate the two kinds of operators. Unlike classical hysterons, the q-hysterons total number is not conserved, i.e. the set of hysterons representing a system is not fixed but obeys probabilistic laws. For a q-hysteron to exist, two probabilistic processes, different in origin, need to take place: a classically thermal activated process followed by quantum tunneling. The quantum origin of the transitions in the q-hysterons yields to operators with unfixed vertical size, unlike the vertically symmetric classical ones, removing the cyclic requirement of the latter (Fig. 2). The vertical size of each transition will depend on which state, on one side of the potential barrier, the quantum magnetization state has tunneled from to the ground state on the other side. Finally, the quantum transitions may not be opposite, but rather along the same direction. For example, after a tunneling event from the left to the right hand side of the double well potential (decreasing transition), a q-hysteron can be created by the still possible tunneling from the left to the right hand side of the potential, but for the magnetic field sweeping in the opposite direction. This new kind of hysteron, called down-down, is schematically described in Fig. 2 (c) and consists of two unidirectional jumps separated by an interval in the applied field $H$. In brief, a q-hysteron is created each time two quantum transitions occur, for magnetic field sweeping in opposite directions.

In this framework, the developed simulation model includes both thermal and tunneling effects. In the present form, the model does not take into account temporal dependence, but it rather calculates the overall magnetization quasi-statically, *i.e.* for a constant field step of typically 1 Oe. For each field step, the thermal activation and tunneling probabilities of $N$ identical and non-interacting molecules (typically $N = 1000$) of spin $S$ are calculated before relaxing to the ground state $\pm S$. The thermal activation probability, $P_{Th}$, gives the probability for the molecule to change from its $S > 0$ ground state to a thermally activated energy level $m_{Th}$:

$$P_{Th}(m_{Th}) = \exp\left(\frac{E(S) - E(m_{Th})}{k_B T}\right) \bigg/ \sum_{m_{Th}=0..\pm S} \exp\left(\frac{E(S) - E(m_{Th})}{k_B T}\right), \qquad (1)$$

where $k_B$ is the Boltzmann constant, $T$ the temperature and the field-dependent ($h_z$) energy of the level $m$ is given by [10]:

$$E(m) = -Dm^2 - g\mu_B h_z m, \qquad (2)$$

where $D$ is the axial zero-field splitting parameter, $g$ is the gyromagnetic factor and $\mu_B$ is the Bohr magneton. For each molecule, the activation occurs if $P_{Th}(m_{Th})$ is higher than the random number between 0 and 1 associated with this molecule. Subsequently, it can either thermally jump over the potential barrier if $E(m_{Th}) \geq 0$, or undergo quantum tunneling, if the energy level difference is less than a fixed value $\Delta E_{Tu}$. In this case, the tunneling probability $P_{Tu}$ between the levels $m$ and $m'$ is given by the Landau-Zener theory [10]:

$$P_{Tu} = 1 - \exp\left(\frac{-\pi\Delta^2}{2\hbar g\mu_B(m'-m)dh_z/dt}\right), \qquad (3)$$

where $\Delta$ is defined as:

$$\Delta = \frac{2D}{((m'-m-1)!)^2}\sqrt{\frac{(S+m')!(S-m)!}{(S-m')!(S+m)!}}\left(\frac{g\mu_B h_x}{2D}\right)^{|m'-m|}. \qquad (4)$$

In summary, the simulation code directly takes into account the concept of q-hysterons, exactly as it was conceived in this work.

For the experimental measurements, we used a $Mn_{12}Ac$ 5 mm elongated single crystal [13]. Magnetization measurements including FORCs, were measured in a commercial physical measurement platform (PPMS - Quantum Design) equipped with a vibrating sample magnetometer insert, with the field along the c-axis. Field sweep rate was kept constant at 50 Oe/s, for temperatures ranging from 2.0 to 4.0 K. On the other hand, simulated major hysteresis curves, as well as FORCs, were obtained for different temperatures ranging from 0 to 10 K. The constants were chosen in order to recreate the $Mn_{12}Ac$ behavior: $S = 10$, $g = 1.9$, $D = 0.399$ cm$^{-1}$ and $h_x = 0.01$ T. The field interval between each first-order reversal curve ($\Delta H_r$) was fixed to 250 Oe for the experimental FORCs, while 500 Oe was used for the simulated results.

The simulation procedure based on our q-hysterons model allows us to quickly reproduce the behavior of a two level quantum system, especially in comparison with the techniques involving matrix diagonalization. The script is able to calculate the 200000 points of a typical magnetization curves at 2 K (between ± 50 kOe, by step of 1 Oe) on a common computer in less than 2 minutes.

Figure 3 shows that the proposed model reproduces the major characteristics of the experimental hysteresis curves. The increase in temperature first promotes the appearance of additional steps, as expected for a thermal activation tunneling process. Beyond a certain temperature value, the steps in the magnetization curve decrease in number and become more rounded in form. Also, the coercive field decreases sensitively until the temperature promotes a purely reversible behavior. It is important to note that these tendencies, while also observed experimentally (see Fig. 1), are not predicted by currently used models. They are explained accepting that the increase in temperature promotes transitions over the potential barrier, without tunneling. This thermally activated process, like in superparamagnetism, results in a decrease of the coercive field.

The FORC technique consists of the successive measurements of minor curves going from a reversal field ($H_r$) to the positive saturation, with $H_r$ values chosen in order to cover the hysteresis area (see insets of Fig. 4). The so-called FORC distribution $\rho$, which experimentally characterizes the complete irreversible behavior of a given system, is given by $H_r$ [11]:

$$\rho(H, H_r) = -\frac{1}{2} \frac{\partial^2 M(H, H_r)}{\partial H \partial H_r} \quad (H > H_r). \tag{5}$$

It is represented as a contour plot in a Preisach plane, where the coercivity axis ($H_c = 0.5(H - H_r)$) and interaction field axis ($H_u = -0.5(H + H_r)$) are directly related to the local irreversible properties, each hysteron having specific $H_c$ (coercivity) and $H_u$ (bias) values (see Fig. 2). Looking as the FORC distribution with the $H_c$, $H_u$ axes, it represents the statistical distribution of irreversible processes (hysterons) related to their local coercivity and bias (or interaction field) values. The FORC method has been successfully used to characterize various systems ranging from geomagnetic samples to antidot arrays [14-28].

Figure 4 exhibits the experimental and simulated FORC diagrams measured at 2 K, while the associated set of QFORCs are presented in the corresponding inset. The similarity between the two results clearly indicates the accuracy of the proposed method of simulation, associating q-hysterons to tunneling process.

The magnetization steps always occur for the same field values, whether on the major or on the QFORCs curves. On the FORC diagrams, they yield a regular network of narrow peaks. Each one of these peaks can be associated to a different q-hysteron of the system.

In both the experimental and simulated cases, the FORC distribution exhibits two distinctive regions, one for positive applied field ($H > 0$) and another for negative applied field ($H < 0$). The first one presents large coercivity ($H_c$) values, but low $H_u$ values. From the QFORCs, one can see that each peak results from the tunneling from negative to positive state (the magnetization increases during the step) after a positive to negative transition. The corresponding q-hysteron shape of this kind of behavior is the down-up type, as shown on Fig. 2 (b). The $H_u$ values can be positive, negative or null, depending if the return transition occurs for applied field values lower, higher or equal than the first transition. The field area covered by the peaks pattern suggests that the system state does not induce a preference among the possible transitions: all positions of the square pattern are occupied by a peak, whose intensity remains similar for both positive and negative $H_u$ values. Hence, after a tunneling process, the populations of each energy level do not remain constant, but always vary accordingly to the Arrhenius thermal activation theory. In most cases, the principal FORC peak arises from a symmetric q-hysteron ($H_u = 0$), suggesting that, for a given temperature, a specific tunneling is favored.

The peak grid of the FORC distribution is perfectly regular in the simulated case, with a field interval between the tunneling processes of 4500 ± 10 Oe, following the expected value of $D/g\mu_B$ [10] [Fig. 4 (b)]. On the experimental result, however, the FORC peak grid is slightly distorted in two different ways [Fig. 4 (a)]. First, as the peaks are aligned horizontally (along the $H$ axis), they are vertically displaced toward lower $H$ values as $|H_r|$ decreases. The regular displacement is around 800 Oe between the peaks associated to $n = 5$ and 3, where $n = |m - m'|$. This vertical distortion, present on all experimental results, can be related to an internal field originated in the dipolar interactions among the $Mn_{12}Ac$ molecules. Also, while the peaks intervals between $n = 3$ and 4, both horizontally and vertically, agree with the expected value, leading to a mean value of 4500 ± 50 Oe, it differs from the $n = 4$ and 5 cases, with a mean value of 4050 ± 50 Oe. In fact, non linear Zeeman splitting with the field could explain this difference. The high degree of precision on the irreversible processes obtained by the FORC results clearly indicates that the tunneling processes involving $n = 5$ happen at lower fields than expected.

For negative applied field, the peaks in the FORC distribution are of low coercivity but high $H_u$ values. Contrary to the peaks located in the $H > 0$ region, they originate from two consecutive magnetization drops, one before and one after the reversal field. The q-hysteron associated to these FORC peaks is therefore of the down-down type [Fig. 2 (c)]. The two tunneling processes involved are separated by a variation of $n$ of only 1 or 2, yielding to the low coercivity observed, as well as the large bias. This behavior is observed both in experimental and simulated results.

One characteristic presented by the experimental FORCs does not appear on the simulated ones: the magnetization smoothly varies reversibly near $H = 0$ (encircled on the Fig. 4 (a) inset), therefore without yielding feature on the FORC distribution. This differs from the simulated curves (see Fig. 3) where, at $H = 0$, it either does not exhibit any step (at low temperature) or shows a step created by tunneling (at high temperature). The effect arises from the internal dipolar transverse fields of the neighboring molecules, allowing quantum tunneling not taken into account in the simulation model.

Finally, the simulation protocol here developed permits one to quickly and easily obtain information that has not yet been experimentally verified. One of the main advantages is that one can observe the full picture concerning the magnetization reversal process that takes place under specific experimental conditions. As an example, Fig. 5 shows the evolution of the percentage of the magnetization reversal, distinguishing between the tunneling and the thermal activation processes. Drawn as function of the applied field, one can see that the tunneling process is clearly the main contribution to the magnetization reversal, effectively occurring for certain field values, as expected. The same information, but drawn as a function of the magnetic quantum number $m$ of the energy level before the tunneling [Fig. 5 (b)], shows that the contribution of the main tunneling process decreases with the temperature, which increases the number of possible initial energy levels. Regarding the thermal magnetization reversal, where $m$ refers to the energy level reached (above the energy barrier), it also tends to decrease with the temperature, the level $m = 0$ being quickly the predominant one as the temperature increases. It is worth noting that inverse magnetization reversal (increasing the magnetization when decreasing the field) occurs for temperatures higher than 1 K, but in very small proportion. The evolution of the tunneling contribution with the temperature is plotted in [Fig. 5 (c)]. After remaining almost 100% until 1.5 K, it decreases until 3.5 K, where it remains to a constant value of around 70%. This relevant information can not be extracted directly from the major hysteresis curves, or any other method. Differently from experiments involving the relaxation time measured from AC susceptibility as a function of temperature (in log scale), where a deviation from a straight line is attributed to a signature of quantum effect [10], in the case of this work, the proposed protocol allows the discrimination between thermal and quantum contributions to the magnetization reversal at a given temperature.

In conclusion, this work developed the QFORC protocol that could be generally applied to explore any quantum systems that present hysteresis described by the QDWP. The QFORC was successfully applied to the special case of the MQT of magnetization of $Mn_{12}Ac$ single molecule magnet. Based on simple and powerful assumptions, as the quantum version of Preisach hysterons, this approach accurately predicts $Mn_{12}Ac$ magnetization reversal, with clear advantage when compared to current models. The very low computational cost involved in the calculations allows one to quickly obtain detailed results, enabling even the simulation of several minor magnetization curves required for the QFORC approach. Additionally, the model predicts several features that were not verified experimentally using standard magnetization measurements, such as the proportion between tunnel and thermal contributions to the magnetization reversal and the identification of the energy levels involved in both processes.

**Acknowledgments**

This work was financially supported by the Brazilian agencies FAPESP, FAPERJ and CNPq.

**Figure captions**

FIG. 1. (Color online) Temperature evolution of experimental major hysteresis curves of a $Mn_{12}Ac$ single crystal (field applied along the magnetization easy axis, $dH/dt$ = 50 Oe/s).

FIG. 2. (Color online) Hysterons (a) classical (b) q-hysteron type down-up (c) q-hysteron type down-down. $H_c$ represents the half-width of the field interval between both transitions, while $H_u$ is its bias.

FIG. 3. (Color online) Temperature evolution of simulated major hysteresis curves of a $Mn_{12}Ac$ single crystal (only one point on each 100 is showed for convenience)

FIG. 4. (Color online) FORC diagram of a $Mn_{12}Ac$ single crystal (field applied along the magnetization easy axis, $dH/dt$ = 50 Oe/s, $T$ = 2 K). Inset: respective QFORCs curves (a) experimental (b) simulated

FIG. 5. (Color online) Evolution of the simulated percentage of magnetization reversal, occurring by tunneling processes (narrow line, open symbols) and by thermal activation (bold lines, solid symbols), with the temperature (a) as function of the applied field going from positive to negative saturation (b) as function of the magnetic quantum number $m$ (c) proportion of thermal and tunneling magnetization reversal as a function of temperature.

Fig. 1, Béron *et al*, Physical Review Letters

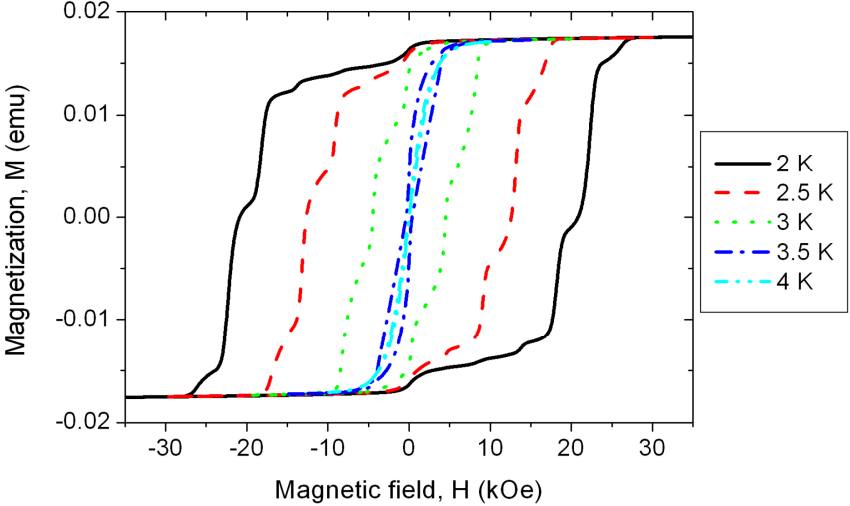

Fig. 2, Béron *et al*, Physical Review Letters

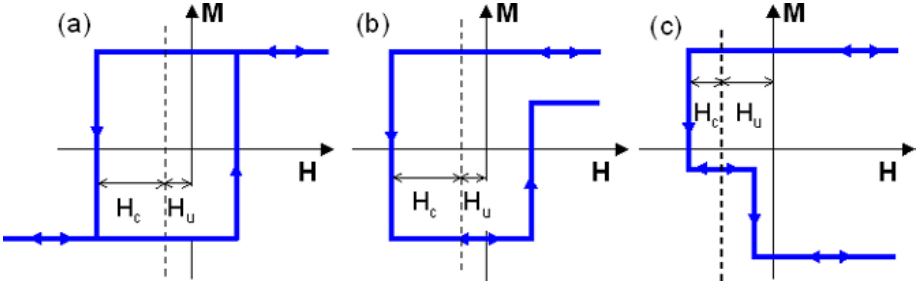

Fig. 3, Béron *et al*, Physical Review Letters

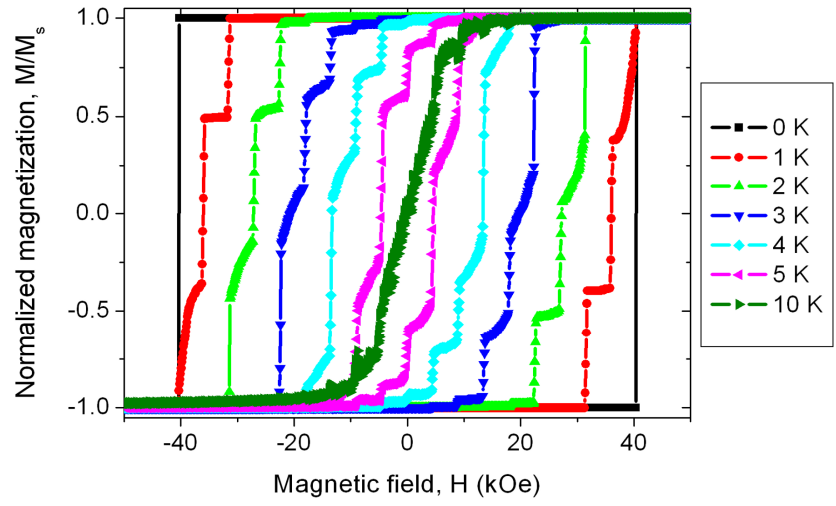

Fig. 4, Béron *et al*, Physical Review Letters

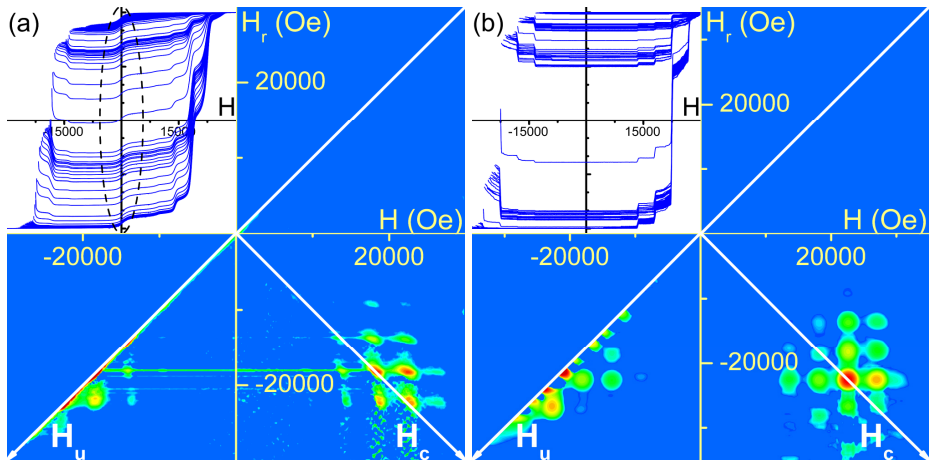

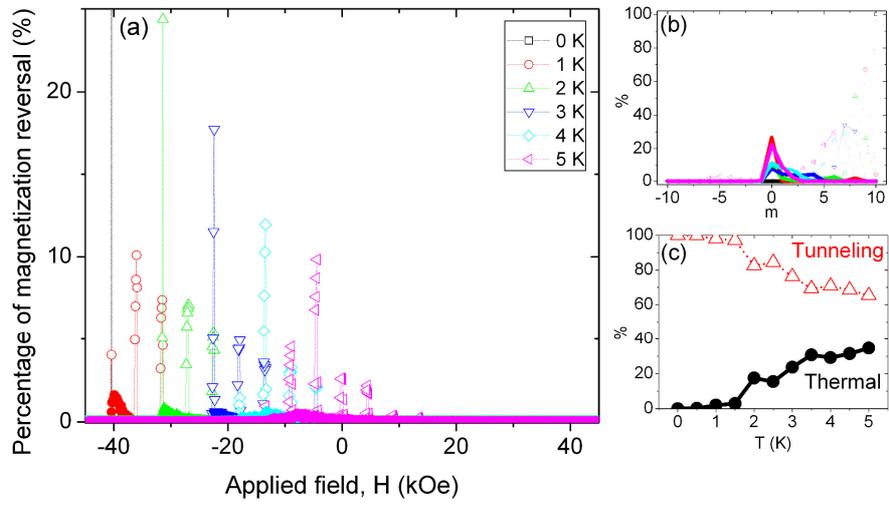